\documentclass{article}
\usepackage{spconf,amsmath,graphicx}
\usepackage{multirow}
\usepackage{dblfloatfix} 
\usepackage{color}
\usepackage{xcolor, colortbl}
\usepackage[most]{tcolorbox} 
\usepackage{booktabs}
\usepackage{multirow}
\usepackage{amsfonts}

\newcommand{\eqcontrib}{$^\dagger$}
\title{EmoShift: Lightweight Activation Steering for Enhanced Emotion-Aware Speech Synthesis}
\name{Li Zhou\eqcontrib\textsuperscript{1}, Hao Jiang\eqcontrib\textsuperscript{1}, Junjie Li\textsuperscript{2}, Tianrui Wang\textsuperscript{3}, Haizhou Li\textsuperscript{1}\sthanks{This work is supported by National Natural Science Foundation of China (Grant No. 62271432), Shenzhen Science and Technology Research Fund (Fundamental Research Key Project, Grant No. JCYJ20220818103001002), Program for Guangdong Introducing Innovative and Entrepreneurial Teams (Grant No. 2023ZT10X044).} \thanks{\eqcontrib Equal contribution. \\Corresponding author: Li Zhou (lizhou21@cuhk.edu.cn)}}
\address{\textsuperscript{1}The Chinese University of Hong Kong, Shenzhen \\
\textsuperscript{2}The Hong Kong Polytechnic University \textsuperscript{3}Tianjin University
}
\begin{document}
\ninept
\maketitle
\begin{abstract}
Achieving precise and controllable emotional expression is crucial for producing natural and context‑appropriate speech in text‑to‑speech (TTS) synthesis. 
However, many emotion‑aware TTS systems, including large language model (LLM)‑based designs, rely on scaling fixed emotion embeddings or external guidance, limiting their ability to model emotion‑specific latent characteristics.
To address this gap, we present \textit{EmoShift}, a lightweight activation‑steering framework incorporating a EmoSteer layer, which learns a steering vector for each target emotion in the output embedding space to capture its latent offset and maintain stable, appropriate expression across utterances and categories. 
With only 10M trainable parameters—less than 1/30 of full fine‑tuning—\textit{EmoShift} outperforms zero‑shot and fully fine‑tuned baselines in objective and subjective evaluations, enhancing emotional expressiveness while preserving naturalness and speaker similarity.
Further analysis confirms the proposed EmoSteer layer’s effectiveness and reveals its potential for controllable emotional intensity in speech synthesis.


\end{abstract}

\vspace{-5pt}
\section{Introduction}
\label{sec:intro}


Emotional expressiveness is a cornerstone of human communication, shaping not only lexical content but also the perceived intent, engagement, and empathy in spoken interaction~\cite{xie2025controllablespeechsynthesisera}. 
In text‑to‑speech (TTS) synthesis, the ability to produce speech with accurate and context‑appropriate emotional qualities is essential for applications such as virtual assistants, audiobooks, and empathetic human–machine dialogue~\cite{10890216, hu-etal-2025-chain}.

Research on emotional control in TTS has made notable progress through explicit modulation of emotion embeddings, often combined with relative attribute learning~\cite{lei2021fine, inoue2024hierarchical}, geometric representations~\cite{cho24_interspeech}, or quantization schemes~\cite{im2022emoq}. 
Other studies leverage generative modeling driven by prosodic cues~\cite{zhang2025proemo} or natural‑language prompts~\cite{du2024cosyvoicescalablemultilingualzeroshot, yang2025emovoice, gao2025emo}.
While effective in varying emotional output, these approaches typically scale a fixed emotion embedding and share all model parameters. This limits interpretability and hinders the direct encoding of emotion‑specific latent dynamics—critical for reliable emotion rendering across both non‑neutral and neutral states.


\begin{figure}[t]
    \centering
    \includegraphics[width=.7\linewidth]{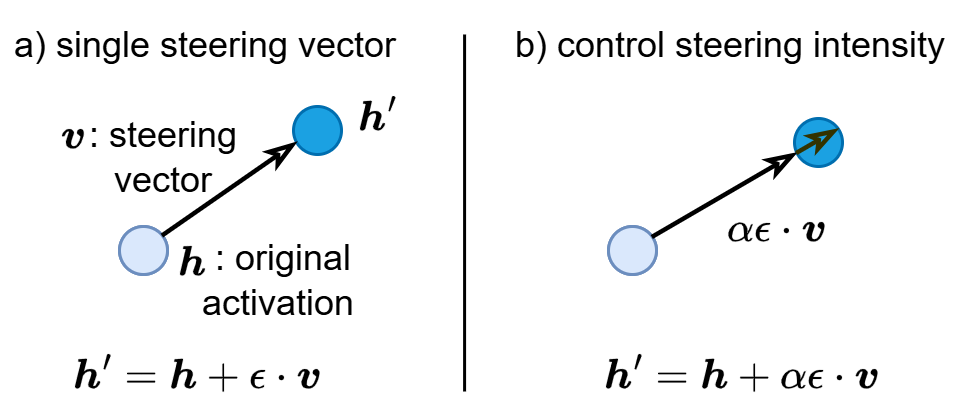}
    \caption{Illustration of steering vector operation in activation space.}
    \label{fig:steering_vector}
\end{figure}

In parallel, activation steering has emerged as a lightweight and interpretable paradigm for controlling generative model behavior~\cite{Dathathri2020Plug, wang2025adaptive}. By learning and injecting targeted directional offsets—steering vectors—into a model’s latent space at inference time (Fig.~\ref{fig:steering_vector}), activation steering enables precise 
control without retraining the core network. It has been applied to style control~\cite{han-etal-2024-word, konen-etal-2024-style}, personalization~\cite{cao2024personalized, bo2025steerablechatbotspersonalizingllms}, safety enforcement~\cite{wang2025adaptive, wu2025automatingsteeringsafemultimodal}, and fairness alignment~\cite{dai2025wordworldevaluatemitigate}.

\begin{figure*}[t]
    \centering
    \includegraphics[width=.95\linewidth]{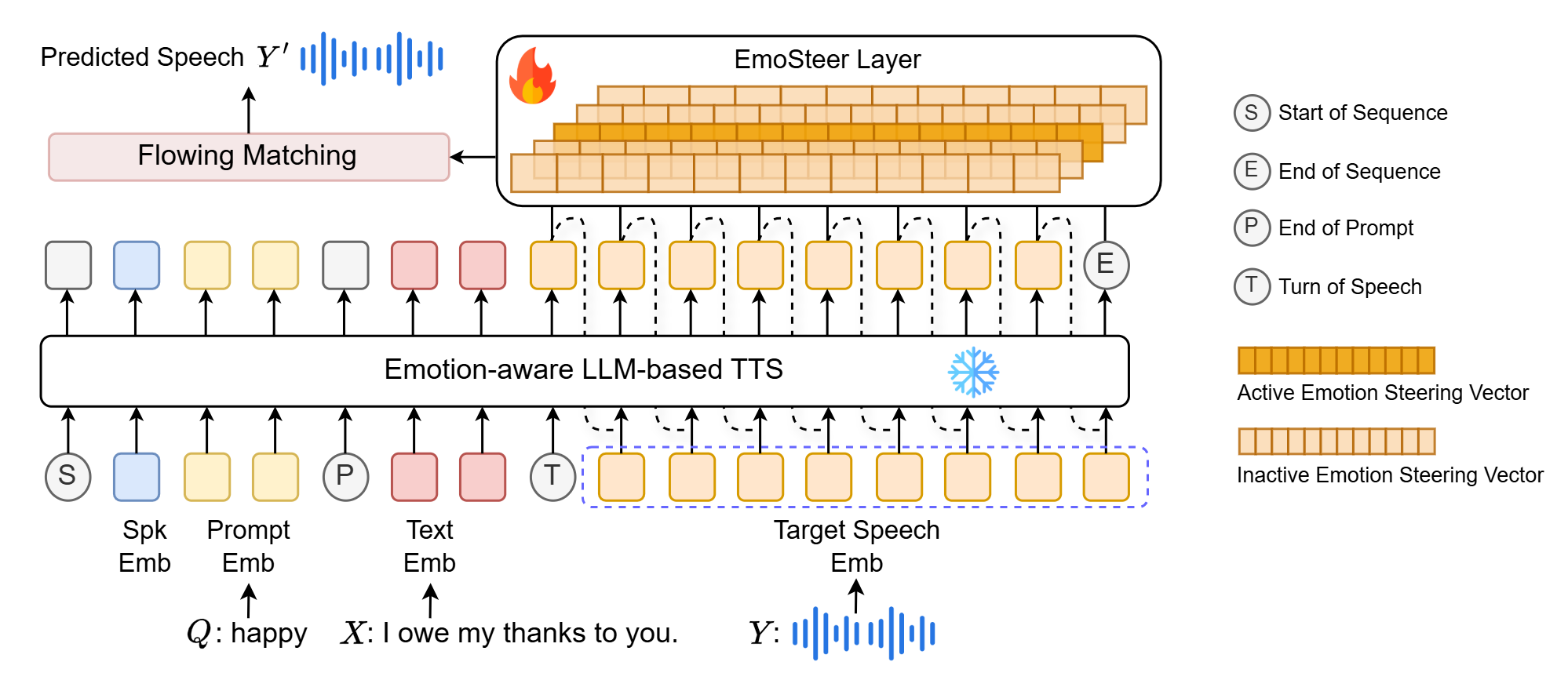}
    \vspace{-0.3cm}
    \caption{Overview of the proposed \textit{EmoShift} framework. Ground-truth speech inputs $Y$ used only during training and omitted during inference.}
    \label{fig:EmoShift}
\end{figure*}

Motivated by the potential of activation steering, we propose \textit{EmoShift}, a lightweight activation‑steering framework for emotion‑aware TTS. At its core is a dedicated EmoSteer layer, which learns emotion‑specific steering vectors in the output embedding space. 
These vectors encode emotion‑dependent latent offsets, enabling precise, consistent, and interpretable emotional control without altering or retraining the base model. 
The design is model‑agnostic and can be seamlessly integrated into LLM‑based TTS pipelines.
Our \textbf{contributions} are as follows:

\begin{itemize} 
\item We propose \textit{EmoShift}, a lightweight activation steering framework for emotion‑aware TTS, incorporating a dedicated EmoSteer layer to learn emotion‑specific steering vectors in the output embedding space for precise and interpretable emotional control. 
\item Extensive objective and subjective evaluations demonstrate that \textit{EmoShift} consistently improves emotional expressiveness over strong emotion‑aware TTS baselines, while preserving naturalness and speaker similarity. 
\item Further analysis reveals that scaling the steering vectors at inference provides an optional mechanism for fine‑grained variation in emotional intensity and expressiveness without compromising emotion‑type fidelity.
\end{itemize}



\section{Related Work}
\label{sec:related_work}

\vspace{-2pt}
\subsection{Emotional Control in Text-to-Speech Synthesis}

Achieving accurate and controllable emotional expression remains a central challenge in expressive speech synthesis.
A common approach performs explicit control by modulating emotion embeddings, where categorical emotion features are injected or scaled to match the target style. Relative attributes frameworks~\cite{lei2021fine,inoue2024hierarchical} learn to distinguish emotions by comparing speech segments to a neutral baseline, and apply the resulting signal at the utterance~\cite{lei2021fine} or phoneme~\cite{inoue2024hierarchical} level. Variants such as EmoQ‑TTS~\cite{im2022emoq} discretize emotion representations, while EmoSphere‑TTS~\cite{cho24_interspeech} encode emotion categories in vector direction for enhanced interpretability.
More recent work leverages generative modeling to guide emotional control. Diffusion‑based systems, e.g., EmoDiff~\cite{guo2023emodiff}, introduce soft‑label guidance between target and neutral emotion distributions, whereas prompt‑driven frameworks~\cite{du2024cosyvoicescalablemultilingualzeroshot,zhang2025proemo,yang2025emovoice} use large language models (LLMs) to interpret natural‑language descriptions (e.g., ``speak happily'') and produce prosody‑controlling signals. 
This line of work offers flexible category control but often with limited precision.

These existing methods either scale a fixed emotion embedding or depend on external guidance. In contrast, our approach learns an emotion‑specific steering vector that directly models a category‑dependent offset in the embedding space, enabling more accurate and consistent emotion rendering in synthesis.

\begin{table*}[ht]
\centering
\scalebox{0.85}{
\begin{tabular}{@{}l|r|rrr|rrrrrr@{}}
\toprule
\multirow{2}{*}{\textbf{Model}}                       & \multicolumn{1}{l|}{\multirow{2}{*}{\textbf{\# Param.}}} & \multicolumn{3}{c|}{\textbf{Speech-Level}}                                                                       & \multicolumn{6}{c}{\textbf{Emotion-Level}}                                                                                   \\ \cmidrule(l){3-11} 
                                                      & \multicolumn{1}{l|}{}                                    & \multicolumn{1}{l}{\textbf{WER $\downarrow$}} & \multicolumn{1}{c}{\textbf{SpkSIM $\uparrow$}} & \multicolumn{1}{c|}{\textbf{DNSMOS $\uparrow$}} & \textbf{Neutral $\uparrow$} & \textbf{Angry $\uparrow$} & \textbf{Happy $\uparrow$} & \textbf{Sad $\uparrow$} & \multicolumn{1}{r|}{\textbf{Surprise $\uparrow$}} & \textbf{Overall $\uparrow$} \\ \midrule
\textbf{CosyVoice}                                    & 0 M                                                      & 7.40                             & 82.23                               & \textbf{3.19}                                 & 74.19            & 86.45          & 61.61          & 61.61        & \multicolumn{1}{r|}{64.52}             & 69.68           \\
\textbf{CosyVoice-SFT}                                & 311 M                                                    & 8.80                             & 92.05                               & 3.16                                 & 70.00            & \textbf{89.68}          & 60.32          & 54.19        & \multicolumn{1}{r|}{74.52}             & 69.74           \\
\textbf{CosyVoice-SFT-Shift}                          & 321 M                                                    & 6.\textbf{80}                             & \textbf{93.03}                               & \textbf{3.19}                                 & 74.84            & 88.39          & \textbf{65.81}          & 62.26        & \multicolumn{1}{r|}{73.23}             & 72.91           \\
\textbf{EmoShift (\textit{Default})} & 10 M                                                     & 7.90                             & 82.41                               & \textbf{3.19}                                 & \textbf{78.39}            & 88.06          & 65.48          & 65.48        & \multicolumn{1}{r|}{73.87}             & 74.26           \\
\textbf{EmoShift (\textit{Best})}    & 10 M                                                     & 11.60                            & 81.50                               & 3.13                                 & 77.10            & 87.10          & 61.61          & \textbf{72.26}        & \multicolumn{1}{r|}{\textbf{81.61}}             & \textbf{75.94}           \\ \bottomrule
\end{tabular}}
\caption{Overall Results: Objective evaluation results of \textit{EmoShift} and baselines in terms of both speech generation quality and emotion generation quality. \textit{Default} uses $\alpha=1$ as in training, whereas \textit{Best} adjusts $\alpha=3$ at inference to strengthen emotion steering.}
\label{tab:overall_obj}
\end{table*}

\vspace{-2pt}
\subsection{Activation Steering Methods for Controlled Applications}
Recent advances in activation steering have shown its potential as a lightweight and interpretable approach for controlling LLMs during inference by injecting well‑designed steering vectors into hidden states, residual streams, or output embeddings. 
Foundational methods such as LM‑Steer~\cite{han-etal-2024-word}, Style Vectors~\cite{konen-etal-2024-style} and In‑context Vectors~\cite{10.5555/3692070.3693379} demonstrated that linear perturbations in latent space can provide continuous and composable control over style, sentiment and task behaviors. 
This paradigm has since been extended to personalization, as in Steerable Chatbots~\cite{bo2025steerablechatbotspersonalizingllms} and Bi‑directional Preference Optimization~\cite{cao2024personalized}; 
to task performance and safety enhancement, including CorrSteer~\cite{cho2025corrsteersteeringimprovestask}, AutoSteer~\cite{wu2025automatingsteeringsafemultimodal} and Contrastive Activation Addition~\cite{rimsky-etal-2024-steering}; 
to knowledge regulation and instruction following, as in SPARE~\cite{zhao-etal-2025-steering} and activation‑based instruction control~\cite{stolfo2025improving}; 
as well as to fairness and cultural alignment through CultureSteer~\cite{dai2025wordworldevaluatemitigate} and to persistent integration via Representation Tuning~\cite{ackerman2024representationtuning}. 
These studies employ diverse vector acquisition strategies and have achieved notable success in text-based and multimodal generation. 

We extend this paradigm to emotion-aware TTS by injecting emotion-specific steering vectors into output embeddings, enabling precise emotional control in generated speech. Our work is closely related to the contemporaneous EmoSteer-TTS~\cite{xie2025emosteer}, which derives emotion-specific activation offsets from few-shot neutral and target activations, whereas EmoShift learns them as trainable steering parameters within the TTS model.




\section{Methodology}

We present \textit{EmoShift}, a lightweight, plug-and-play activation steering framework for emotion-aware LLM-based TTS (Fig.~\ref{fig:EmoShift}). The proposed system first models TTS as a conditional auto-regressive speech token generation problem within the LLM framework, and then integrates an emotion-specific activation steering layer to achieve explicit and controllable emotional expression.

\vspace{-2pt}
\subsection{LLM-based Problem Formulation and Modeling Setup} 
\label{sec:method-formulation}
We formulate emotion-aware TTS as a conditional auto-regressive token generation task. The model is conditioned on three sources of information: 
a speaker embedding $\boldsymbol{s} \in \mathbb{R}^d$, 
an emotion prompt $Q$ (e.g., ``happy'') represented as a sequence of prompt embeddings $\{\boldsymbol{q}_i\}_{i=1}^u$, and text embeddings $\{\boldsymbol{x}_j\}_{j=1}^n$ derived from the speech script $X$. The LLM then generates a sequence of discrete speech tokens $\{\boldsymbol{y}'_k\}_{k=1}^m$ until a special end-of-sequence token $\textcircled{E}$ is predicted. These tokens are subsequently decoded into a speech waveform $Y'$ using a flow matching-based vocoder.


To encode all conditioning information, the input sequence is organized with special tokens marking key boundaries:
\begin{equation}
\left[
  \text{\textcircled{S}},\ 
  \boldsymbol{s},\ 
  \{\boldsymbol{q}_i\}_{i=1}^u,\ 
  \text{\textcircled{P}},\ 
  \{\boldsymbol{x}_j\}_{j=1}^n,\ 
  \text{\textcircled{T}},\ 
  \{\boldsymbol{y}_k\}_{k=1}^m,\ 
  \text{\textcircled{E}}
\right],
\end{equation}
where $\textcircled{S}$, $\textcircled{P}$, $\textcircled{T}$, and $\textcircled{E}$ denote start of sequence, end of prompt, turn of speech, and end of sequence, respectively. 
During training, ground-truth tokens $\{\boldsymbol{y}_k\}$ are included with teacher-forcing scheme; 
at inference, generation starts after $\textcircled{T}$ and proceeds autoregressively without ground-truth context.
The inference distribution is:
\begin{equation}
P(\boldsymbol{y}'_{1:m}) 
= \prod_{k=1}^m p(\boldsymbol{y}'_{k} \mid \boldsymbol{s}, \{\boldsymbol{q}_i\}, \{\boldsymbol{x}_j\}, \boldsymbol{y}'_{<k}),
\end{equation}
and the model is trained by minimizing the negative log-likelihood of ground-truth tokens:
\begin{equation}
    \mathcal{L} =-\sum_{k=1}^m{\log p\left( \boldsymbol{y}_k \right)}.
\end{equation}
\vspace{-2pt}
\subsection{\textit{EmoShift}: Emotion-Specific Activation Steering} 
While emotion-aware LLM-based TTS supports multi-condition generation, it does not make the emotion expression space explicit in the model’s parameters, limiting interpretability and direct control.
To address this, we introduce \textit{EmoShift}, a lightweight activation steering layer that learns explicit emotion-specific directions in the latent space and applies them in a plug-and-play manner. By projecting each hidden state $\boldsymbol{h}$ originally generated from emotion-aware LLM-based TTS onto emotion-dependent offsets, \textit{EmoShift} makes the emotion representation space interpretable, manipulable, and controllable, without changing the core architecture or retraining the base LLM-based TTS model.

Given a target emotion $e$ from the prompt $Q$, \textit{EmoShift} computes a steering vector $\boldsymbol{v}_e = \boldsymbol{h} \mathbf{W}_e$ for each hidden state $\boldsymbol{h} \in \mathbb{R}^d$ using a learnable projection matrix $\mathbf{W}_e \in \mathbb{R}^{d\times d}$. The hidden state is then modified as:
\begin{equation}
\boldsymbol{h}' = \boldsymbol{h} + \epsilon \cdot \boldsymbol{v}_e,
\end{equation}
where $\epsilon$ is a fixed base scaling factor controlling adjustment magnitude during training. Here, $\mathbf{W}_e$ captures the emotion-specific activation shift pattern, and $\boldsymbol{v}_e$ encodes expressive deviations from neutral prosody associated with emotion $e$.
During inference, a gain factor $\alpha \geq 1$ is introduced for fine-grained control over emotional intensity:
\begin{equation}
\boldsymbol{h}' = \boldsymbol{h} + \alpha\epsilon \cdot \boldsymbol{v}_e.
\end{equation}
By adjusting $\alpha$, the emotional expression can be smoothly modulated while preserving the target emotion identity.
All steering directions are stored in compact projection matrices $\mathbf{W}_e$, adding minimal parameter overhead. Once trained, \textit{EmoShift} can be seamlessly integrated into existing LLM-based TTS pipelines without architectural changes or retraining, enabling interpretable and portable emotion rendering.

\section{Experiments}
\vspace{-2pt}
\subsection{Experimental Setup}
\textbf{Dataset:} 
We use the English subset of the ESD dataset~\cite{zhou2021emotional}, consisting of 350 parallel utterances produced by 10 English speakers under 5 emotional states (neutral, happy, angry, sad, and surprise). Following the official data partitioning, we allocate 300, 20, and 30 utterances to the training, development, and test sets, respectively, while ensuring that all speaker–emotion variants of each utterance are assigned to the same set.
\textbf{Training details: }
We adopt CosyVoice-300M-Instruct (hereafter CosyVoice)~\cite{du2024cosyvoicescalablemultilingualzeroshot} as the emotion-aware LLM-based TTS backbone, into which the proposed EmoSteer layer is inserted to construct our \textit{EmoShift} model. For \textit{EmoShift}, we introduce five learnable emotion steering vectors (including neutral as one of the emotions) based on the dataset. During training, we set the steering coefficient $\epsilon=0.001$ by default, adopt a learning rate of $1\times10^{-4}$, and train the model for 5 epochs.
\textbf{Baselines: } For comparison, we consider three baselines: 1) the backbone CosyVoice itself, 2) CosyVoice-SFT, a fully fine-tuned version of CosyVoice, and 3) CosyVoice-SFT-Shift, where the EmoSteer layer is inserted and fine-tuned on top of the fully fine-tuned CosyVoice.

\vspace{-2pt}

\subsection{Evaluation Metrics}

We employ both objective and subjective evaluations to compare the proposed \textit{EmoShift} model with the baselines. For \textbf{objective evaluation}, speech generation quality is assessed using the Word Error Rate (WER), computed by applying the Whisper-Large-v3 ASR model~\cite{radford2023robust} to the synthesized speech, the Speaker Similarity (SpkSIM), measured with the WavLM-Base model, and the overall perceived quality, evaluated using DNSMOS~\cite{reddy2021dnsmos} for each synthesized utterance. Emotion generation quality is evaluated using the emotion2vec model~\cite{ma-etal-2024-emotion2vec}, which perform speech emotion recognition (SER) on the generated audio to obtain the classification accuracy for each emotion type.  A higher classification accuracy indicates a stronger ability of the TTS model to generate emotion-aware synthetic speech.
For \textbf{subjective evaluation}, we consider the Mean Opinion Score (MOS), which rates the overall audio quality and naturalness on a five-point scale ranging from 1 (bad) to 5 (excellent), and the Emotion Mean Opinion Score (Emo-MOS), which rates the perceived emotional expressiveness on the same scale. For Emo-MOS, the focus is solely on whether the model has learned to produce speech with the intended emotion, without considering the closeness of the generated expression to that of the ground truth.

\vspace{-2pt}
\subsection{Experimental Results}

As shown in Table~\ref{tab:overall_obj}, \textit{EmoShift} surpasses its backbone across all emotion categories, demonstrating the effectiveness of the proposed EmoSteer layer in enhancing emotional expressiveness.
With only 10M trainable parameters—less than 1/30 of a fully fine-tuned system—\textit{EmoShift} not only outperforms the fully fine-tuned CosyVoice-SFT in both overall emotion classification accuracy and per-emotion performance (except for the Angry category), but also achieves comparable overall results to CosyVoice-SFT-Shift, which combines full-parameter fine-tuning with the EmoSteer layer.
Increasing $\alpha$ from 1 (\textit{Default}) to 3 (\textit{Best}) boosts emotion-aware TTS performance by amplifying emotion-specific offset subspaces, achieving 75.94\% overall recall and notable gains for \textit{Sad} and \textit{Surprise}.
At the same time, \textit{EmoShift} maintains speech generation quality at a level close to that of its backbone and fully fine-tuned counterparts, with differences remaining within an acceptable range.

\begin{table}[t]
\centering
\begin{tabular}{@{}lrr@{}}
\toprule
\textbf{Model}         & \textbf{MOS} & \textbf{Emo-MOS} \\ \midrule
CosyVoice     & 4.07±0.10    & 3.67±0.14            \\
CosyVoice-SFT & 3.93±0.13    & 3.79±0.14            \\
EmoShift      & \textbf{4.14±0.09}    & \textbf{3.96±0.12}            \\ \bottomrule
\end{tabular}
\caption{Subjective evaluation scores obtained from 10 listeners, with 95\% confidence intervals estimated using the t-test.}
\label{tab:subj_mos}
\end{table}

We also conduct a human subjective evaluation with 10 listeners, where each listener rates 25 utterances with balanced emotions (5 samples per emotion) for CosyVoice, CosyVoice-SFT, and \textit{EmoShift}. Each score is reported with its 95\% confidence interval (CI95), computed using the t-distribution over listener ratings.
As shown in Table~\ref{tab:subj_mos}, \textit{EmoShift} attains the highest MOS (4.14) and Emo-MOS (3.96), further confirming its improvements in both naturalness and emotional expressiveness over the baselines.


\begin{table}[t]
\centering
\begin{tabular}{l l r}
\toprule
\textbf{Base Model} & \textbf{Metric} & \textbf{Win Rate (\%)} \\
\midrule
\multirow{2}{*}{CosyVoice}     & MOS   & 71.95 \\
                               & Emo-MOS  & 80.65 \\
\multirow{2}{*}{CosyVoice-SFT} & MOS   & 72.00 \\
                               & Emo-MOS  & 80.30 \\
\bottomrule
\end{tabular}
\caption{
Win rates (\%) of versions with the \textit{EmoSteer} layer over their base models, reported for MOS and Emo-MOS.}
\label{tab:emos_layer_winrate}
\end{table}

\section{Analysis}
\vspace{-2pt}
\subsection{Effectiveness of the EmoSteer Layer}
To verify the contribution of the proposed EmoSteer layer, we conduct a human evaluation using two base models—CosyVoice and CosyVoice-SFT—where the presence or absence of the EmoSteer layer serves as the model variable. 
10 listeners participate in preference tests on the MOS and Emo-MOS dimensions, selecting the better sample for each pair. Table~\ref{tab:emos_layer_winrate} reports the win rates (\%) of the versions with the EmoSteer layer over their respective base models. 
When applied to CosyVoice, the EmoSteer layer achieves win rates of 71.95\% for MOS and 80.65\% for Emo-MOS; when applied to the fully fine-tuned CosyVoice-SFT, the corresponding win rates are 72.00\% and 80.30\%. 
These consistently high win rates across both metrics and base models demonstrate that the EmoSteer layer effectively enhances emotional expressiveness while maintaining, and in some cases improving, speech naturalness, which is consistent with the objective evaluation results in Table~\ref{tab:overall_obj}.

\begin{figure}[t]
    \centering
    \includegraphics[width=0.8\linewidth]{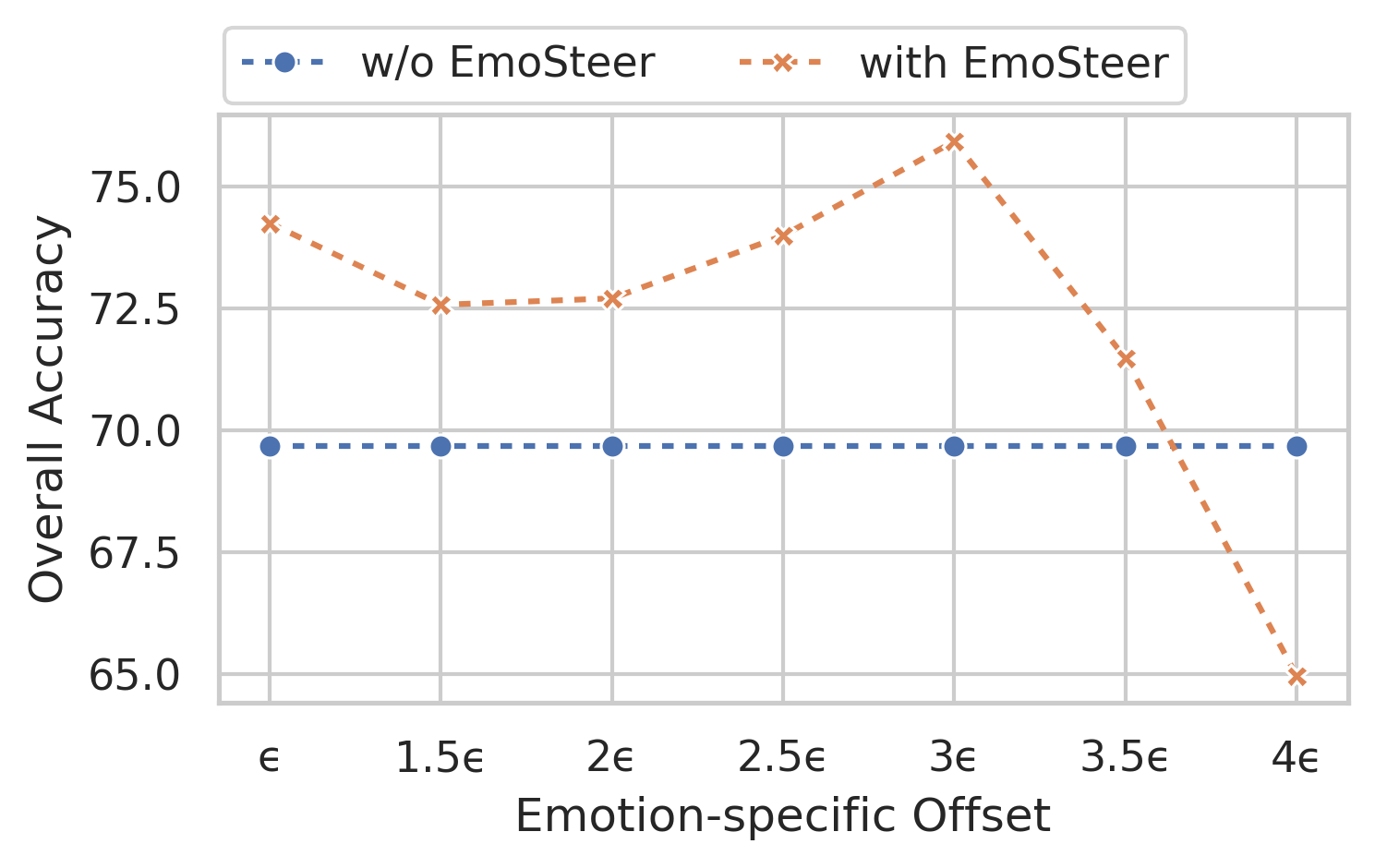}
    \caption{Overall emotion recognition accuracy with different steering factor $\alpha$ at inference.}
    \label{fig:steering_offset}
\end{figure}

\begin{table}[t]
\begin{tabular}{@{}lrrrrr@{}}
\toprule
\textbf{Emotion}       & \textbf{Neutral} & \textbf{Angry} & \textbf{Happy} & \textbf{Sad} & \textbf{Surprise} \\ \midrule
\textbf{Win Rate (\%)} & 55.84            & 64.48          & 48.55          & 61.24        & 68.39             \\ \bottomrule
\end{tabular}
\caption{Per-emotion win rates (\%) in AB preference tests on emotional intensity when increasing the steering factor $\alpha$ from $1$ to $3$.}
\label{tab:ab_pref_results}
\end{table}

\vspace{-2pt}
\subsection{Impact of Steering Coefficient at Inference}

We hypothesize that increasing the steering factor value $\alpha$ at inference enlarges the learned emotion-specific offset subspaces, thereby strengthening the emotional expression in synthesized speech and making it more likely to be correctly recognized by the emotion2vec-based SER model.
To test this hypothesis, we scale $\alpha$ from its training value $1$ up to $4$ in increments of $0.5$ on the test set, and measure the overall emotion recognition accuracy using the same SER model. 
As shown in Fig.~\ref{fig:steering_offset}, 
models with EmoSteer outperform those without it for all coefficient settings except $\alpha=4$. 
Except for the training default $\alpha=1$, the accuracy increases steadily from $1.5\epsilon$ to $3\epsilon$ where it peaks, before dropping sharply at $4\epsilon$. 
A similar pattern is observed in LLM text generation~\cite{han-etal-2024-word}, where increasing the steering value from negative to positive reduces both the number and intensity of toxic words.

To assess perceptual effects, we conduct AB preference tests on emotional intensity with 10 listeners. For each emotion, five utterances are randomly sampled. For each utterance, the steering factor is varied from $\alpha=1$ to $3$ in increments of $0.5$, generating five versions in sequence. 
Adjacent versions are paired and presented to listeners, who are asked to choose the one with the stronger perceived emotional expression. 
The win rate is then calculated for the version with the larger steering factor.
As shown in Table~\ref{tab:ab_pref_results}, four of the five emotions achieve win rates above 50\%, with Surprise (68.39\%) and Angry (64.48\%) showing the largest gains, followed by Sad (61.24\%) and Neutral (55.84\%).
These results indicate that a moderate increase in $\alpha$ can strengthen the expressive intensity of most emotions with minimal impact on speech quality, whereas excessive increase may cause degradation.

\section{Conclusion}

In this work, we introduce \textit{EmoShift}, a lightweight activation‑steering framework for emotion‑aware TTS built around a dedicated EmoSteer layer that learns emotion‑specific steering vectors in the output embedding space. 
This design enables precise and interpretable emotional control in a parameter‑efficient and architecture‑agnostic manner. 
Extensive evaluations indicate that \textit{EmoShift} delivers consistently stronger emotional expressiveness than competitive baselines while maintaining naturalness and speaker similarity, and further analysis highlights its optional capability for fine‑grained adjustment of emotional intensity without compromising emotion‑type fidelity.
Future work will extend \textit{EmoShift} to more emotional categories, including compound emotions, and develop adaptive steering strategies for richer expressive control.


\vfill\pagebreak

\footnotesize
\bibliographystyle{IEEEbib}
\bibliography{refs}

\end{document}